\documentclass[runningheads]{llncs}
\usepackage{graphicx}
\usepackage{svg}

\usepackage{tikz}
\usepackage{comment}
\usepackage{amsmath,amssymb} 
\usepackage{color}

\usepackage[accsupp]{axessibility}  

\usepackage{hyperref}

\usepackage[width=122mm,left=12mm,paperwidth=146mm,height=193mm,top=12mm,paperheight=217mm]{geometry}

\begin{document}
\pagestyle{headings}
\mainmatter
\def\ECCVSubNumber{15}  

\title{Representation Learning for Non-Melanoma Skin Cancer using a Latent Autoencoder} 

\titlerunning{Representation Learning for Non-Melanoma Skin Cancer} 

%
\author{Simon M. Thomas\inst{1}\orcidID{0000-0003-4609-2732}}

\authorrunning{S. M. Thomas}
%
\institute{
Australian e-Health Research Centre, CSIRO, Herston, Queensland, Australia.
}
\maketitle


\begin{abstract}

Generative learning is a powerful tool for representation learning, and shows particular promise for problems in biomedical imaging. However, in this context, sampling from the distribution is secondary to finding representations of real images, which often come with labels and explicitly represent the content and quality of the target distribution. It remains difficult to faithfully reconstruct images from generative models, particularly those as complex as histological images. In this work, two existing methods (autoencoders and adversarial latent autoencoders) are combined in attempt to improve our ability to encode and decode real images of non-melanoma skin cancer, specifically intra-epidermal carcinoma (IEC). Utilising a dataset of high-quality images of IEC ($256 \times 256$), this work assesses the result of both image reconstruction quality and representation learning. It is shown that adversarial training can improve baseline FID scores from 76 to 50, and that benchmarks on representation learning can be improved by up to 3\%. Smooth and realistic interpolations of the variation in the morphological structure are also presented for the first time, positioning representation learning as a promising direction in the context of computational pathology.

\end{abstract}

\section{Introduction}

Generative learning is well-established as a method to learn meaningful representations of 
diverse and complex image distributions \cite{brock2018large}. Importantly, the quality of the representation is made explicit by the quality of the models output, where its structure can be explored by exhaustive sampling, or systematic interrogation using targeted queries or data labels to perform attribute editing \cite{pidhorskyi2020adversarial}. Furthermore, treating the knowledge within the model as a hypothesis, visualization techniques provide us with a way to understand what the model has learned and whether it corresponds to our own understanding of the world. Excitingly, it may also help us to learn new structure within the data we had not otherwise known. Such ability has enormous appeal to the biomedical imaging community, in particular, helping to understand the complex problem of the histological diagnosis of disease.

There have been several applications of generative models to histological images of various diseases, including skin cancer \cite{thomas2021characterization}, breast cancer \cite{quiros2019pathologygan}\cite{karras2020training} and ovarian cancer \cite{levine2020synthesis}. However, instead of just being able to generate synthetic images that imitate features of cancer, 
it is desirable to generate \emph{real} images. In this way, we possess explicit ground truth of good quality images and relevant features, and so are better situated to assess the quality of generated images and the underlying representation.

There are several methods that perform representation learning by jointly learning to encode and decode real images \cite{donahue2019large}\cite{xia2022gan}. A notable approach is that of adversarial latent autoencoders (ALAEs) \cite{pidhorskyi2020adversarial}, which produces compelling, but not exact reconstructions of images at scales of $28 \times 28$, $128 \times 128$, $256 \times 256$ and $512 \times 512$ pixels \cite{pidhorskyi2020adversarial}\cite{thomas2021characterization}. This architecture was originally designed to be trained in a progressively grown manner. However, recent techniques in training GANS suggest that autoencoder components can improve the quality and stability of generative modelling \cite{liu2020towards}. That is, a system which can learn to reconstruct images complements the task of generative representation learning. This raises the question of whether it is possible to combine these two methods to improve our ability to reconstruct real images faithfully.

This work explores this question in the context of non-melanoma skin cancer, specifically intra-epidermal carcinoma (IEC). IEC is a common form of skin cancer which is interesting due to it presenting in terms of degrees of dysplasia i.e varying and increasing degrees of cellular disorder \cite{patterson2014weedon}. This makes it particularly well-suited to generative learning methods which attempt to represent variation within images in a continuous manner. It utilizes the high-quality IEC data introduced by \cite{2022thomasHighlyExpressiveMachine}, which additionally provides a useful point of comparison in terms of reconstruction quality and representation learning.

\section{Methods}

All associated code and data can be found at \url{https://github.com/smthomas-sci/RepresentationLearningNon-Melanoma}. The networks were implemented using Tensorflow v2.5 and trained on 2 x 1080GTX 8GB NVIDIA GPUs. 

\subsection{Intraepidermal Carcinoma Dataset}

\begin{figure}[h!]
\centering
\includegraphics[height=6.5cm]{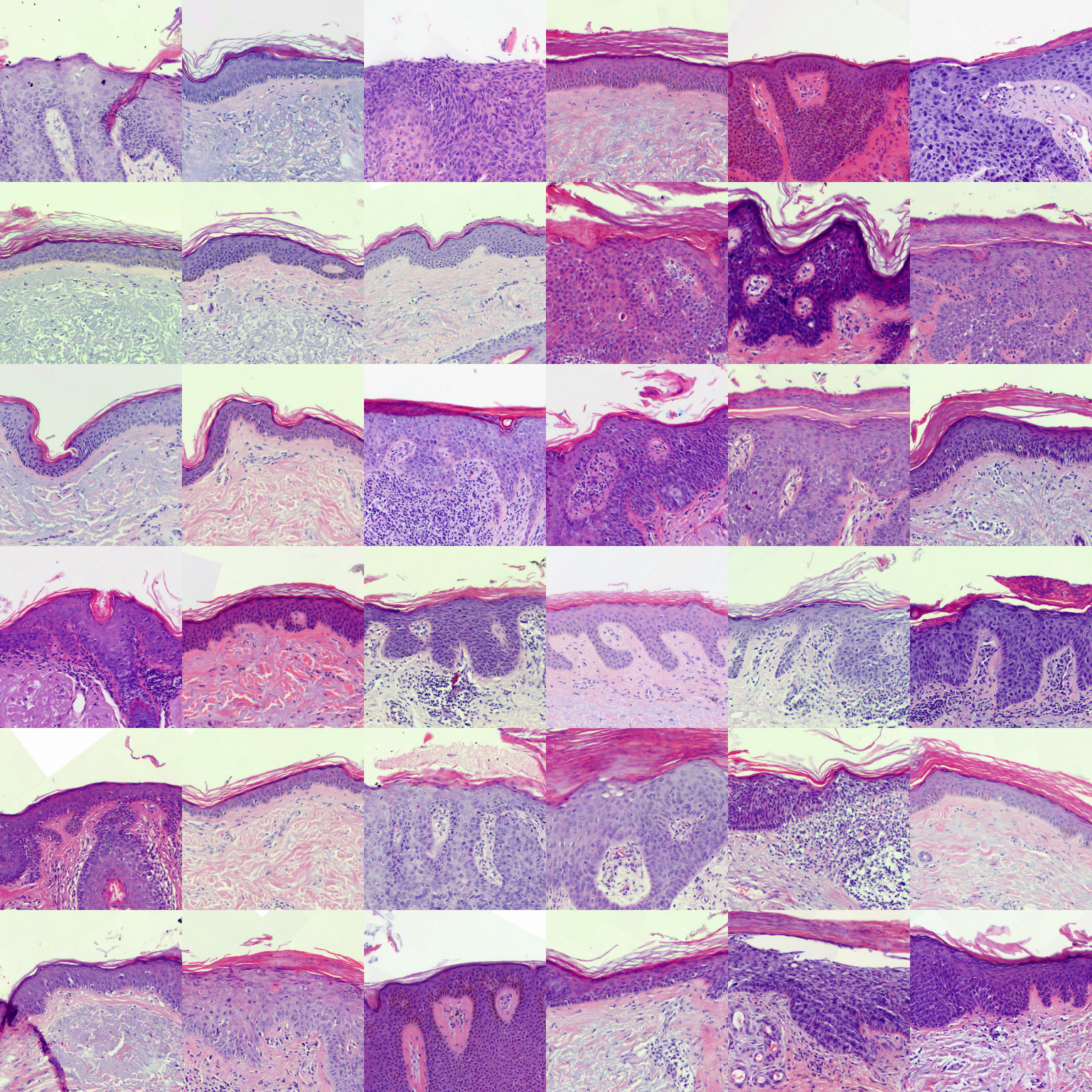}
\caption{Example images of IEC at resolution $ 256 \times 256$ pixels. There is variation in background color and staining color, as well as the full variety of tissue morphology found in the keratin, epidermal and dermal regions of the image.}
\label{fig:1}
\end{figure}

This work utilises the dataset introduced by \cite{2022thomasHighlyExpressiveMachine}, which consists of 16,980 images of intraepidermal carcinoma skin cancer at $256 \times 256$ pixel resolution. Each image is aligned to  provide a fixed and consistent perspective of the epidermis, which is where non-melanoma skin cancers propagate from \cite{patterson2014weedon}. Either side of the epidermis is the keratin layer (superior) and the dermal layer (inferior), which although not strictly indicative of a particular cancer, exhibit morphological changes correlated with disease e.g. inflammation. The images represent variations of tissue morphology within the keratin, epidermal and papillary dermal layers, as well as physical variations such as orientation ink, staining and lighting. Examples of the variety are shown in Fig.~\ref{fig:1}.

Associated with the dataset are natural language captions which characterizes each image systematically using a controlled vocabulary. Following \cite{2022thomasHighlyExpressiveMachine}, this study utilizes the simplified labels such that the three layers can be classified into the following classes: 

\begin{itemize}
  \item \textbf{Keratin}: thin / thick and basketweave keratosis / parakeratosis
  \item \textbf{Epidermis}: mild / full-thickness dysplasia
  \item \textbf{Dermis}: normal dermis / solar damaged / inflammation
\end{itemize}

\subsection{Network Architecture}

\begin{figure}[h!]
\includegraphics[height=8.5cm]{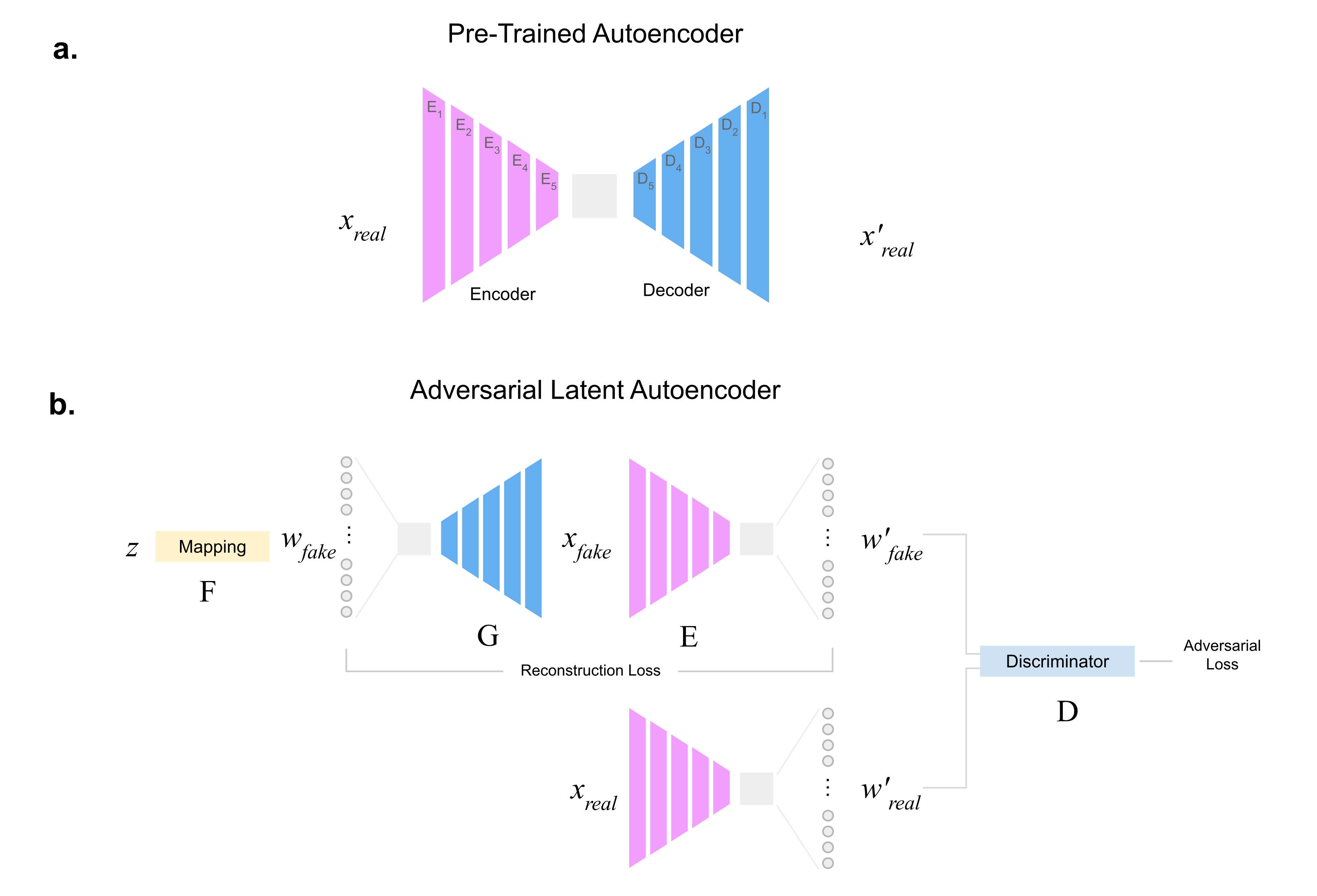}
\caption{ \textbf{a.} An autoencoder network composed of an encoder and decoder. Each component is further divided into 5 stages of down sample and up sample residual blocks. The autoencoder is trained greedily one stage at a time to reconstruct the input image $x$ using mean absolute error. All spatial information is retained through trained. \textbf{b.} The encoder and decoder networks are reconfigured to form an adversarial latent autoencoder (ALAE) network. This network is composed of a mapping network $F$ which maps random samples from $z$ to a learned latent space $w$. This feeds a generator network $G$, part of which includes the decoder from before. The images is then encoded with the $E$ network, which includes the encoder, mapping the input image back to the $w$ latent space. The dimensionality of $w$ is 512. $w$ vectors for both real and fake images are fed to a discriminator $D$ for adversarial training. A reconstruction loss is also included to encourage 1:1 mapping between latent vectors and images.}
\label{fig:2}
\end{figure}

The network training procedure was decomposed into 2 phases. Phase 1 involved an autoencoder network (Fig.~\ref{fig:2}a) which learned to reconstruct the input, but retained the spatial dimensions.  The autoencoder network (denoted as encoder and decoder) was trained in stages, where each stage corresponds to the addition of a residual block. Each block was followed by either an up or down sample operation, and channel expansion via $1 \times 1$ convolutions. The residual layers consisted of 2 lots of $3\times3$ convolutions with zero-padding, batch normalization, leaky ReLU (slope of 0.2), with a skip connection between the input and final features. The number of filters for each block were: $\{256 \times 256: 32, 128 \times 128: 64, 64 \times 64: 128, 32 \times 32: 256, 16 \times 16: 512, 8 \times 8: 512\}$.

Each stage was greedily trained using mean absolute error, with all previous layers being locked. For example, the first stage consisted of an input image $x$ being reconstructed using, $D_1(E_1(x)) = x\prime$, and the second stage as $\text{D}_1(\text{D}_2(\text{E}_2(\text{E}_1(x))))  = x\prime$, where $E_1$ and $D_1$ were locked. This was done for 5 stages (Fig.~\ref{fig:2}a). The ADAM optimizer with a learning rate of 0.001 was used across 10 to 30 epochs for each stage with a batch size of 32, and then finally the whole network was fine-tuned over 60 epochs with a learning rate of 0.0001.

Phase 2 reconfigured the encoder and decoder network into an adversarial latent autoencoder (Fig.~\ref{fig:2}b). The encoder, outputting an $8 \times 8 \times 512$ feature block, was flattened and fully-connected to a $512$-dimensional latent space, $w$, constituting the $E$ component. The $G$ component performed the inverse, feeding back to the decoder to produce image $x\prime$. Samples were taken from 
$ z \sim \mathcal{N}(\mu=0, \sigma^{2}=1)$ and then mapped to $w$ using a 3-layer fully-connected mapping network $F$ with Leaky ReLU activations (slope=0.2). The discriminator $D$ was also 3-layered fully-connected network with ReLU activations, transforming $w$ features into a real or fake prediction logit. Following the original ALAE\cite{pidhorskyi2020adversarial}, a non-saturating adversarial loss and L2 reconstruction loss was used. The original encoder layers of $E$ and decoder layers of $G$ were locked, and whole network was trained for 120 epochs using the ADAM optimizer with a learning rate of 0.0001 and a batch size of 8. The exponential moving average of the $F$ and $G$ weights were used for all visualizations.

\subsection{Learned Concept Vectors}

For a fair comparison with \cite{2022thomasHighlyExpressiveMachine}, logistic regression models where trained on latent space features $w$ for the following binary classification tasks.

\begin{itemize}
  \item thin keratin vs. thick keratin
  \item basketweave keratosis vs. parakeratosis
  \item mild dysplasia vs. full-thickness dysplasia
  \item normal dermis vs. solar damage
  \item normal dermis vs. inflammation
\end{itemize}

The concept vectors were then defined using the normalized coefficients of the logistic regression models, 
where the weights describe the direction of orthogonal separation between the classes. The vectors were then used to perform latent space exploration and attribute editing.

\section{Results \& Discussion} 

\subsection{Image Reconstructions \& Sampling}

The quality of the reconstructions for both the autoenoder and latent autoencoder networks was compared using the FID score \cite{heusel2017gans}. The base-line reconstructions $x \prime$ compared to their input $x$, had an FID score of 77.8. This was improved under adversarial training, with the latent autoencoder samples and reconstructions scoring an FID of 50.3 and 50.0 respectively. Notably, the reconstruction quality improved despite the encoder and decoder weights being locked. However, the original VQ-GAN model \cite{2022thomasHighlyExpressiveMachine} achieved an FID score of 17.6, and 36.1 using a latent space projection, indicating a substantial difference in realism.

\begin{figure}[h!]
\includegraphics[height=6.5cm]{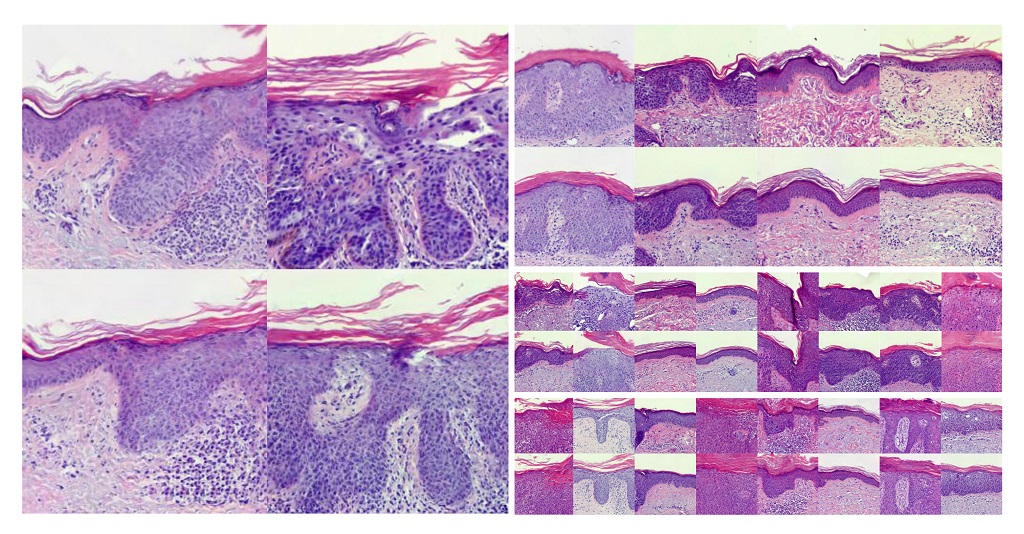}
\caption{Reconstructions of $256 \times 256$ pixel IEC images from the unseen test set. Global structure and tissue features are successfully modelled across a large variety of image contexts, in particular the keratin and dermal layers. Notably, important cellular features related to IEC cancer are not realistically modelled and instead show repeated textures with cell-like appearance. Zoom in for more details.}
\label{fig:3}
\end{figure}

\begin{figure}[h!]
\centering
\includegraphics[height=6.5cm]{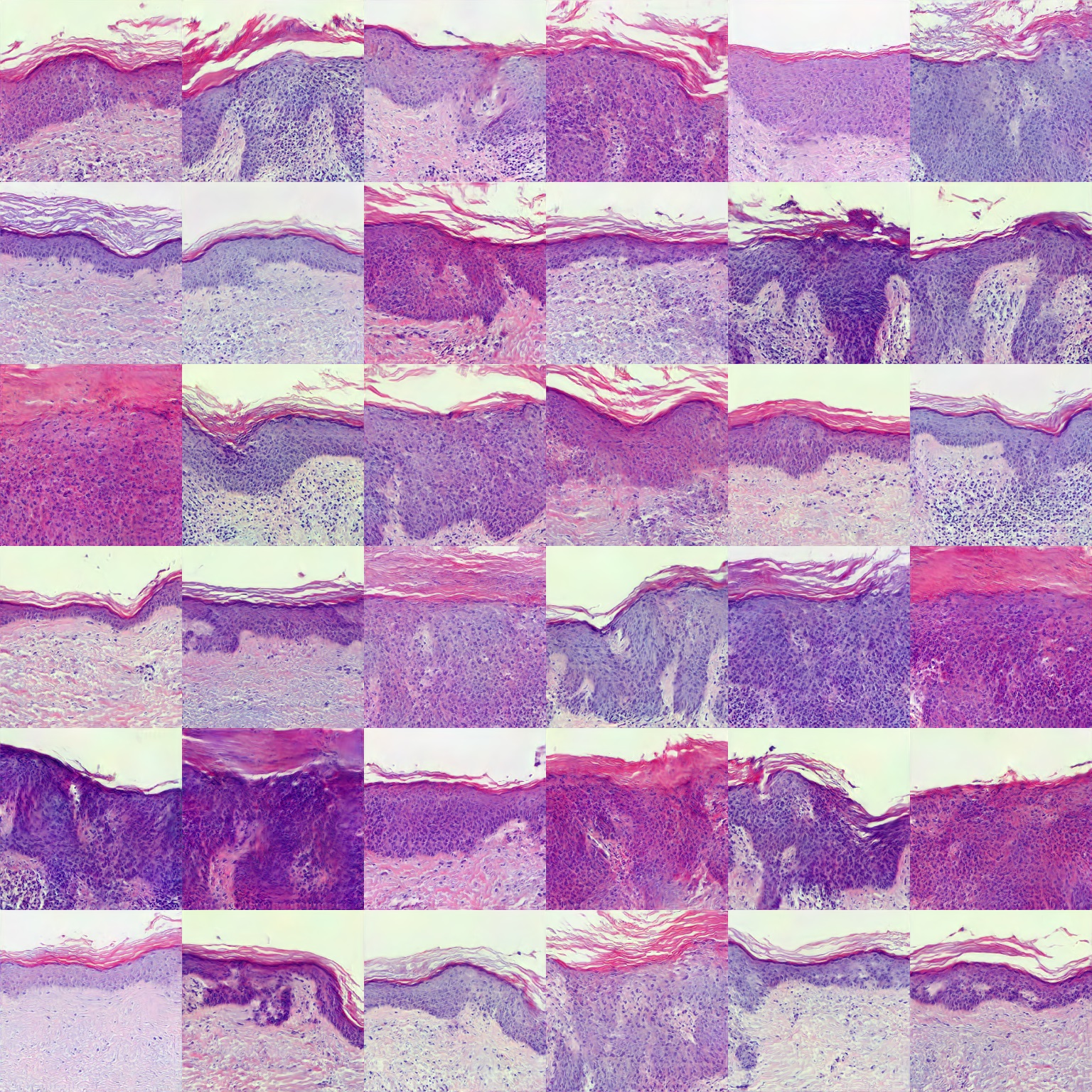}
\caption{Random samples of $256 \times 256$ pixel IEC images. As seen in reconstructions, global structure and tissue features are successfully modelled across a large variety of image contexts. Sample quality matches that of reconstruction quality i.e. FID 50.0. Zoom in for details. The continuity and structure of the latent space can be further seen in the \href{https://drive.google.com/file/d/1ij1TMekV6f0RcngUb9C_JFwEtfW06ANh/view?usp=sharing}{[accompanying video]}}
\label{fig:4}
\end{figure}

Visually, the latent autoencoder reconstructions capture the global structure of the images quite well (Fig.~\ref{fig:3}). It is clear that a large amount of the variation in terms of tissue types, morphological change, lighting and staining conditions are able to be learned by the model. This is further demonstrated in Fig.~\ref{fig:4}, which shows random samples from the latent space. In both the reconstructions and samples, the keratin and dermal features more closely resemble realistic features. However, it is clear that the epithelial tissue, that is, the features most relevant to cancer, are not faithfully reproduced. It appears that high-frequency textures that show basic cellular features are produced instead. This was observed by \cite{2022thomasHighlyExpressiveMachine} as well, reiterating the challenges of learning highly variable cellular features using a reconstruction objective. Other work with GANs indicates that such features can be learned for cancer images \cite{quiros2019pathologygan}, and with realistic but non-exact reconstructions \cite{quiros2020learning}\cite{thomas2021characterization}. Thus, successfully representing hiqh-resolution real data remains the goal.

Importantly, projecting images into a continuous space where each point ideally represents a realistic image, enables smooth interpolations showing the transformation of various tissue types - \href{https://drive.google.com/file/d/1ij1TMekV6f0RcngUb9C_JFwEtfW06ANh/view?usp=sharing}{[see accompanying video]}. As so far known, these visualisations demonstrate for the first time meaningful transitions between tissue types and the correlated morphological changes in the context of skin cancer. Importantly, is provides initial evidence that image generation techniques can be utilized to model patterns, and perhaps eventually discover new ones, that characterize the continuous nature of cancer progression. Of course, what the model learns is itself only a hypothesis, and must be verified experimentally.

The fact that the adversarial training improved the baseline reconstruction quality suggested that further gains could be made by including an adversarial term in the autoencoder formulation, similar to the VQ-GAN. However, after testing this idea with a PatchGAN during pretraining, the resulting latent autoencoder performed no better i.e. FID 53. This suggests that the limitation is not in the encoder and decoder network, but instead results from the lack of expressivity in the latent space to capture finer details such as epithelial cells. I suspect then that much of the models representative power is instead utilised for generating diverse reconstructions in terms of global structure. One possible direction to overcome this would be crop the images to remove some of the variation.

Compared to the original ALAE training regime, it was found that the pre-training of the encoder and decoder significantly improved the stability of the adversarial training. This was observed reliably in early experiments with the MNIST, CIFAR-10 and CelebA datasets at resolutions between $28\times 28$ and $64 \times 64$ pixels (examples can be seen on the \href{https://github.com/smthomas-sci/RepresentationLearningNon-Melanoma}{[code repository]}). This hints at a possible explanation for the effectiveness of transfer learning with GANs, where the tasks is to adapt the latent space to the new distribution, more so than updating the convolution layers. It again indicates that the better the autoencoder components, the better the quality of the latent autoencoder results. Of course, as discovered, this is only up to a point, where the capacity of the latent space may actually be the limitation. This is an interesting conjecture to pursue in future research.

\subsection{Representation Quality \& Attribute Manipulation}

The binary classification task for the 5 concept vectors revealed an overall improvement of between 1-3\% for all concepts compared to \cite{2022thomasHighlyExpressiveMachine}. Indeed, the performance across the training, validation, and test sets (Table ~\ref{table:table1}), suggest that the quality of the representation improved overall, showing more consistent and generalised performance, with a closer match between the three data splits compared to \cite{2022thomasHighlyExpressiveMachine}. This could be a result of training the network to project images into a structured and smooth latent space, demonstrated by the capability to sample images (Fig. ~\ref{fig:4}). That is, the direct enforcement of continuity between locations in the latent space may have conferred more generalised performance in a way a more simple compression objective cannot.

\setlength{\tabcolsep}{4pt}
\begin{table}[h!]
\begin{center}
\caption{Concept vector binary classification scores produced from a logistic regression classifier
trained to predict positive (+) and negative (-) concept classes. The classifier takes
the $w$ representation of the image, where $w$ is a vector of size 512. The members of each class for each dataset is the same as \cite{2022thomasHighlyExpressiveMachine}. }
\label{table:table1}
\begin{tabular}{lcllll}
\hline\noalign{\smallskip}
Concept & Binary Classes & Dataset & Accuracy & Sensitivity & Specificity \\
\noalign{\smallskip}
\hline
\noalign{\smallskip}
& & & & &  \\
Keratin    & Thin            & Train   & 0.9813   & 0.9846   & 0.9785 \\
           & vs.             & Val.    & 0.9840   & 0.9848   & 0.9833 \\
Thickness  & Thick Keratosis & Test    & 0.9853   & 0.9973   & 0.9753 \\
& &                                    & +2.45\%  & +1.35\%  & +3.36\%        \\


& & & & &  \\
    & Basketweave            & Train   & 0.8618   & 0.8788   & 0.8367 \\
Parakeratosis & vs.          & Val.    & 0.8522   & 0.8588   & 0.8424 \\
  & Parakeratosis            & Test    & 0.8411   & 0.8592   & 0.8104  \\
& &                                    & +3.05\%  & +2.06\%  & +4.74\%        \\


& & & & &  \\
    & Mild Dysplasia            & Train   & 0.9805   & 0.9786   & 0.9817 \\
Dysplasia & vs.                 & Val.    & 0.9424   & 0.9320   & 0.9502 \\
  & Full-thickness              & Test    & 0.9519   & 0.9375   & 0.9602 \\
& &                                       & +1.97\%  & +3.04\%  & +1.36\%        \\


& & & & &  \\
    & Normal Dermis            & Train   & 0.8336   & 0.7076   & 0.9024 \\
Solar Damage & vs.             & Val.    & 0.8347   & 0.5802   & 0.9211 \\
  & Solar Damage               & Test    & 0.8352   & 0.6059   & 0.9215 \\
& &                                      & +1.29\%  & -10.0\%  & +5.53\%        \\


& & & & &  \\
    & Normal Dermis            & Train   & 0.8841   & 0.8152   & 0.9260 \\
Inflammation & vs.             & Val.    & 0.8289   & 0.6203   & 0.9572 \\
  & Inflammation               & Test    & 0.8516   & 0.6382   & 0.9526 \\
& &                                      & +2.46\%  & -0.89\%  & +4.03\%        \\

\hline
\end{tabular}
\end{center}
\end{table}
\setlength{\tabcolsep}{1.4pt}

\begin{figure}[h!]
\centering
\includegraphics[width=10.5cm]{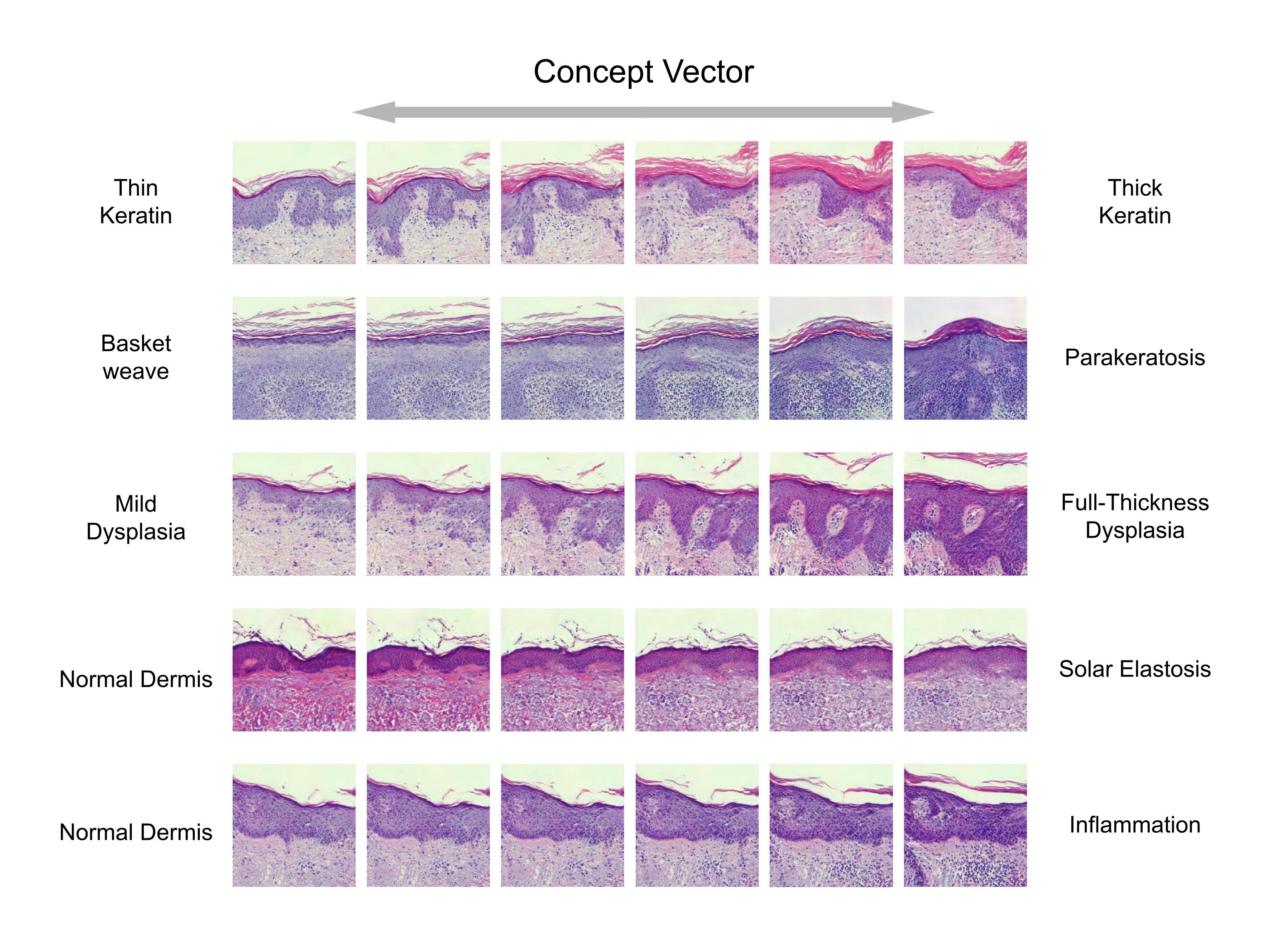}
\caption{ Interpolations for learned concept vectors related to the keratin, epidermal and dermal layers of the skin. The image transformations indicate that concepts transition smoothly, with correlated features
changing and independent features showing hardly any variation. An interactive version of this figure 
can be explored on the \href{https://smthomas-sci.github.io/RepresentationLearningNon-Melanoma}{GitHub page})}
\label{fig:5}
\end{figure}

The manipulation of images according to the various concept vectors is shown in Fig. ~\ref{fig:5}, which further indicates that the latent space is structured in a meaningful way. Indeed, the transitions show much more realistic journeys through the latent space compared to \cite{2022thomasHighlyExpressiveMachine}. This is  especially true for the dysplasia concept, which shows the natural progression of increased disorder and thickening of the epidermis. Similarly, solar damage and inflammation can be modified without substantial changes to the epidermal and keratin layers. The thin and thick keratin concept appears to be slightly entangled with epidermal features, but the healthy dermis remains conceptually the same throughout the transition. In contrast, there is a strong association with parakeratosis and dysplasia, which is seen in the second row of Fig. \ref{fig:5}, where the composition of the image substantially changes. In this case, the concepts are entangled with a variety of features, which may or may not correspond to biological reality. In practice, it is desirable for the concepts to be disentangled, such as seen in face manipulations \cite{pidhorskyi2020adversarial}. Of course, this remains a considerable challenge, for histological images of skin cancer are a much more complex interaction of features compared to human faces. 

In general, image manipulations via concepts are a powerful way to explore what a model has learned. Interestingly, concepts defined in terms of simple binary categories may be too coarse to describe the nuances of cellular morphology. In the long term, we can expect to be able learn and transverse semantic representations of language to explore a joint latent space, such as recently demonstrated with interpolations and manipulations using a CLIP and image generation model \cite{ramesh2022hierarchical}. A prototype example of this is demonstrated on the \href{https://smthomas-sci.github.io/RepresentationLearningNon-Melanoma}{GitHub page}), serving as an early step in that direction for histopathology.

\section{Conclusion}

It is shown that generative modelling can be a powerful tool for learning diverse and biologically meaningful representations of histological images of skin cancer. A desirable result is the ability to faithfully represent real data, requiring a system to both encode and decode (generate) real data. To that end, architectural choices were made to prioritise reconstructing real images, while also attempting to represent the features in a continuous and structured latent space via adversarial learning. The result is a model which can successfully represent broad image features across a large variety of tissues types, orientations, lighting and staining conditions. Furthermore, the relationships between the images can be visualized by traversing the latent space, demonstrated for the first time for non-melanoma skin cancer images. Systematic manipulation of images using concept vectors provides further insight into the quality of the learned representation, and in general constitutes an improvement by comparison to other similar work in the field. This work identifies several promising areas where further progress can be made and paints a picture of the longer term goals in the application of representation learning to problems in computational pathology.

\clearpage

%
%
\bibliographystyle{splncs04}
\bibliography{egbib}

\end{document}